\documentclass[secnumarabic,twocolumn,aps,prl,showpacs,showkey,superscriptaddress,prl,floatfix]{revtex4-1}

\usepackage{amsmath,amssymb,amsthm}

\usepackage{hyperref}
\usepackage{amsthm}
\usepackage{graphicx}
\usepackage{subfigure}
\usepackage{color,bbm}
\usepackage[normalem]{ulem}

\definecolor{violet}{rgb}{1.00,0.00,1.00}	

\graphicspath{{./Figures/}}

\newcommand{\R}{\mathbb{R}}

\newcommand{\x}{\mathbf{x}}

\newcommand{\m}{\mathbf{m}}
\newcommand{\y}{\mathbf{y}}

\begin{document}

\title{Synchronization in random balanced networks}

\author{Luis Carlos Garc\'ia del Molino}%
\email[]{garciadelmolino@ijm.univ-paris-diderot.fr}
\author{Khashayar Pakdaman}
\affiliation{Institut Jacques Monod, CNRS UMR 7592, Universit\'e Paris Diderot, Paris Cit\'e Sorbonne, F-750205, Paris, France}
\author{Jonathan Touboul}
\affiliation{Mathematical Neuroscience Team, CIRB-Coll\`ege de France and INRIA Paris-Rocquencourt, BANG Laboratory, 11 place Marcelin Berthelot, 75005 Paris, France}
\author{Gilles Wainrib}
\affiliation{Laboratoire Analyse G\'eom\'etrie et Applications, University Paris 13, France}

\date{\today}%

\begin{abstract}
Characterizing the influence of network properties on the global emerging behavior of interacting elements constitutes a central question in many areas, from physical to social sciences. 
In this article we study a primary model of disordered neuronal networks with excitatory-inhibitory structure and balance constraints.
We show how the interplay between structure and disorder in the connectivity leads to a universal transition from trivial to synchronized stationary or periodic states. 
This transition cannot be explained only through the analysis of the spectral density of the connectivity matrix.
We provide a low dimensional approximation that shows the role of both the structure and disorder in the dynamics.
\end{abstract}

\maketitle

\section{Introduction}

Networks representing interactions in physical, biological or social systems exhibit a structured connectivity and, at the same time, a high degree of disorder~\cite{complex-networks}.  In this work, we show how the interplay between these two leads to novel forms of synchrony that could not exist if either one is suppressed.

Neuronal networks are a paramount example of structured connectivity with a high degree of disorder. On the one hand, they are highly structured. Very often a neuron is either excitatory or inhibitory, a principle known as Dale's law. Moreover, beyond such physiological constraints, balanced networks (i.e.~networks where the excitatory and inhibitory input to a given cell balance each other) are currently  a major subject of study \cite{shadlen-newsome:94,brunel:00, vanvreeswijk-sompolinsky:96}. On the other hand, neuronal networks display characteristic disorder properties in the interconnection strengths~\cite{parker:03,marder-goaillard:06}. Studies show that certain levels of disorder, rather than being detrimental, might be functional~\cite{aradi-soltesz:02,santhakumar2004plasticity,soltesz2005diversity}.





In this article we present a detailed analysis of the behavior of excitatory-inhibitory balanced neural networks with random synaptic weights, and report a novel transition related to the level of disorder that cannot be explained only through the properties of the spectral density. We investigate the dynamics of a canonical model of random neural network~\cite{sompolinsky-crisanti-etal:88,amari:72}:\\
\vspace{-0.5cm}
\begin{equation}\label{eq:GeneralNetwork}
	\dot{x}_i=-x_i+\sum_{j=1}^n J_{ij} S(x_j)
\end{equation}
with random synaptic coefficients $J_{ij}$. In this model, $x_i$ represents the activity of neuron $i$, $S(\cdot)$ is a sigmoid function accounting for the synaptic response, and $J_{ij}$ corresponds to the synaptic weight from neuron $j$ onto neuron $i$. 
%
The coefficients we will consider in this article are defined as $J_{ij}=\mu m_j + \sigma \xi_{ij}$ with $\mathbf{m}=(m_i)_{1\leq i\leq n}$ a normalized vector with sum zero corresponding to the structure  of the network (e.g.~excitatory/inhibitory), and $(\xi_{ij})_{1\leq i,j\leq n}$ centered weakly correlated random variables with variance $\chi_j^2/n$ satisfying the \emph{balance condition} $\sum_j \xi_{ij}=0$ (This balance condition can be relaxed, the analysis remains valid for coefficients such that $\sum_{i}\xi_{ij}=o(1)$ as $n\to \infty$). $\mu$ and $\sigma$ are two scalars that control the presence of structure and disorder in the connectivity.
In~\cite{sompolinsky-crisanti-etal:88}, the authors studied system~\eqref{eq:GeneralNetwork} with $\mu=0$ and independent, non-balanced Gaussian coefficients with $\chi_j=1$. They discovered in the large $n$ limit a phase transition at a critical value of the disorder $\sigma$ between a trivial state where all trajectories converge to $0$ and a chaotic regime centered around $0$, which was further analyzed in~\cite{cessac:94,wainrib-touboul:13}. In contrast to these studies, the case of $\mu\neq 0$ has been only partially explored. 
In~\cite{rajan-abbott:06}, the authors investigated the spectral properties of matrices $J= \sigma \xi + \mu M$ as defined above. They proved that when $\xi$ is balanced, the average synaptic weight $\mu \m$ has no impact on the spectrum of $J$, which is identical to that of $\sigma \xi$ (for which they computed the limit distribution $n\rightarrow \infty$ when $\xi$ is Gaussian). From numerical observations, these authors also reported that in the non-balanced case, while the bulk of the spectrum of $J$ is distributed as in the balanced case, there are a few eigenvalues, referred to as outliers, that deviate significantly from it. Mathematical characterization of these spectral properties for general finite rank perturbations $\m$ of random independent identically distributed matrices was done in \cite{tao2011outliers}. None of these previous studies dwell with the full dynamics of \eqref{eq:GeneralNetwork}, which is the topic of the present paper.
Our motivation stems from the fact that balanced networks have the property that the net mean-field input vanishes, and therefore complex dynamics are likely to emerge from specific patterns of the \emph{fluctuations} around the mean activity. This mechanism is different from the usual mean-field theory~\cite{hermann-touboul:12}. We now show that indeed, balanced networks tend to display a surprising regularity with highly synchronized activity. 

\begin{figure*}[t]
 \centering
 \hspace{-0.5cm}
 \includegraphics[width=0.33\textwidth]{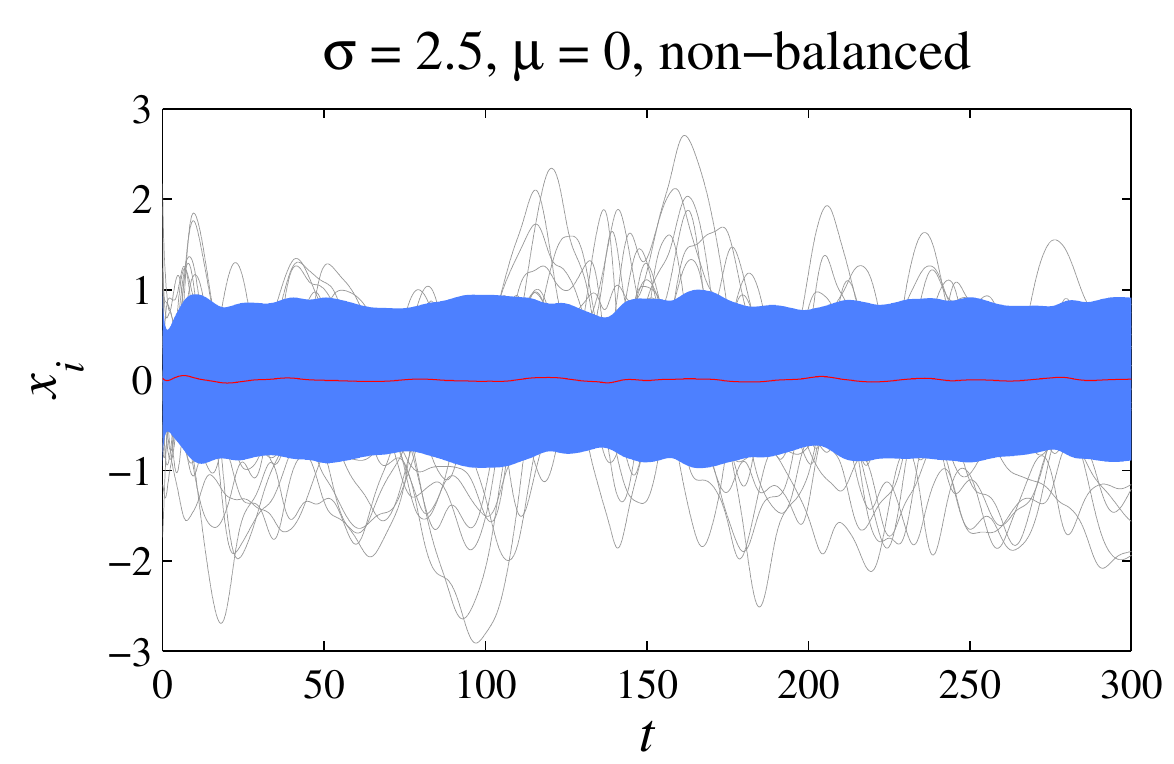}
 \includegraphics[width=0.33\textwidth]{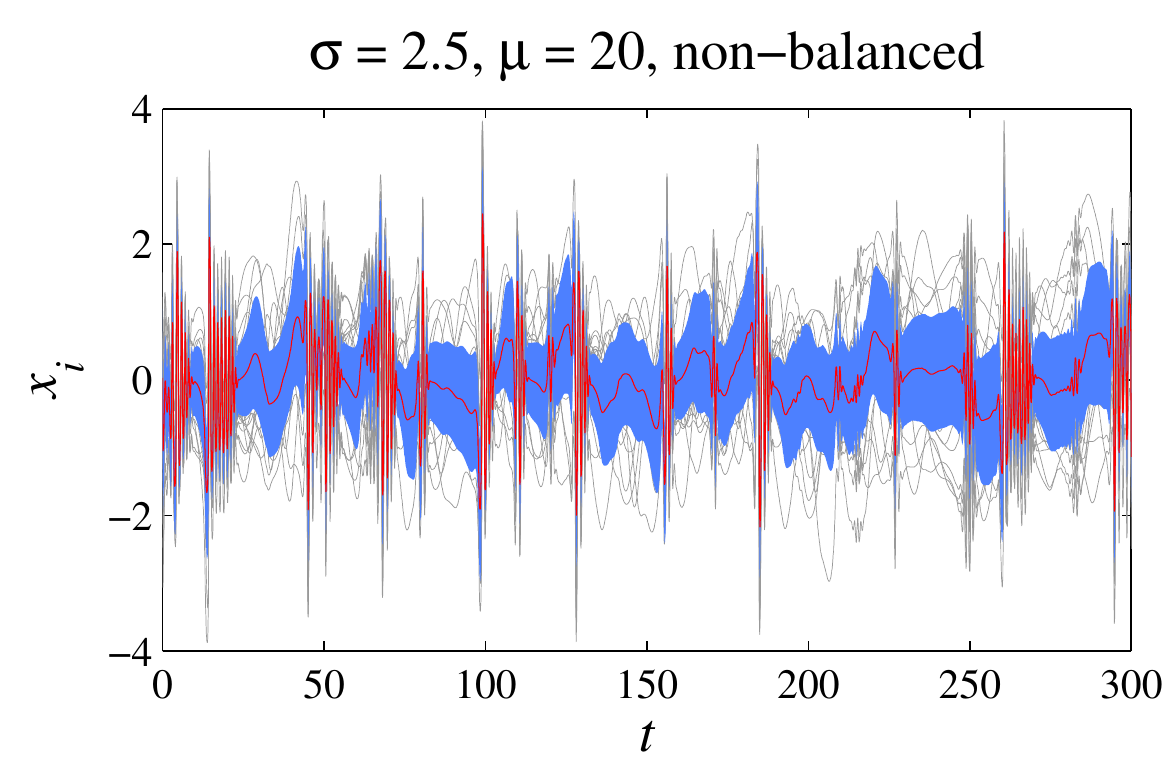}
 \includegraphics[width=0.33\textwidth]{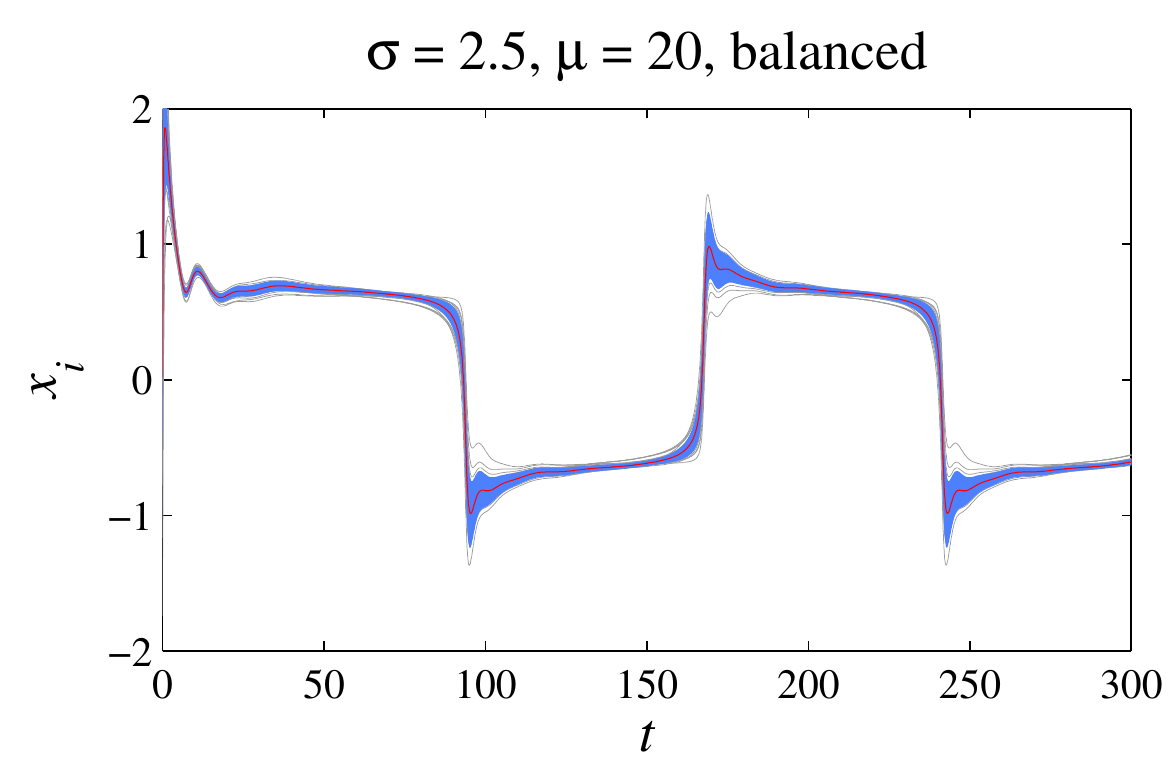}\\
 \hspace{-0.5cm}
 \includegraphics[width=0.33\textwidth]{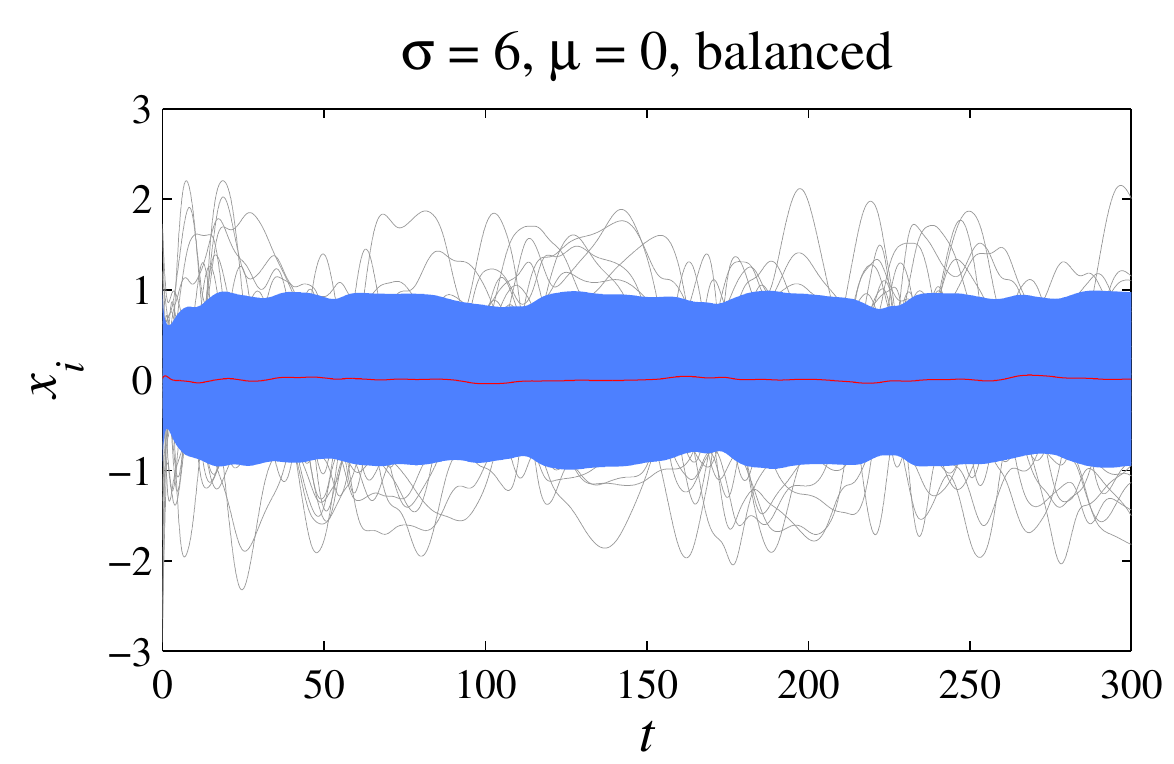}
 \includegraphics[width=0.33\textwidth]{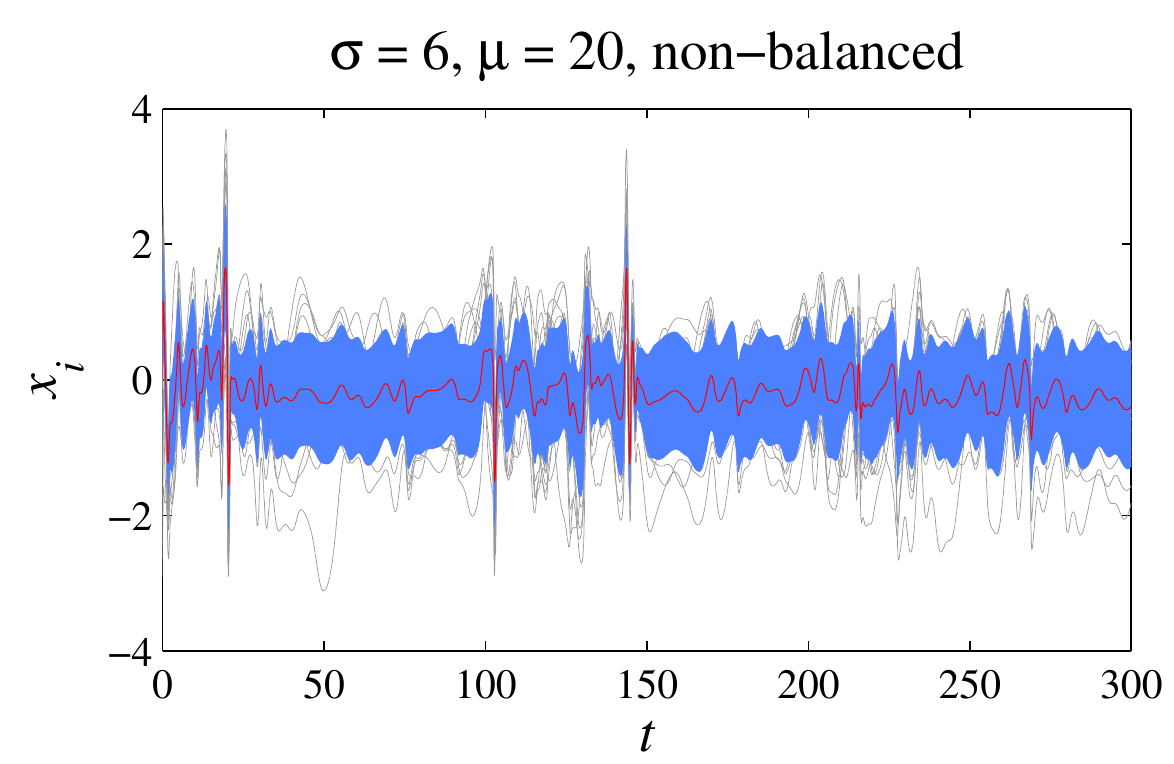}
 \includegraphics[width=0.33\textwidth]{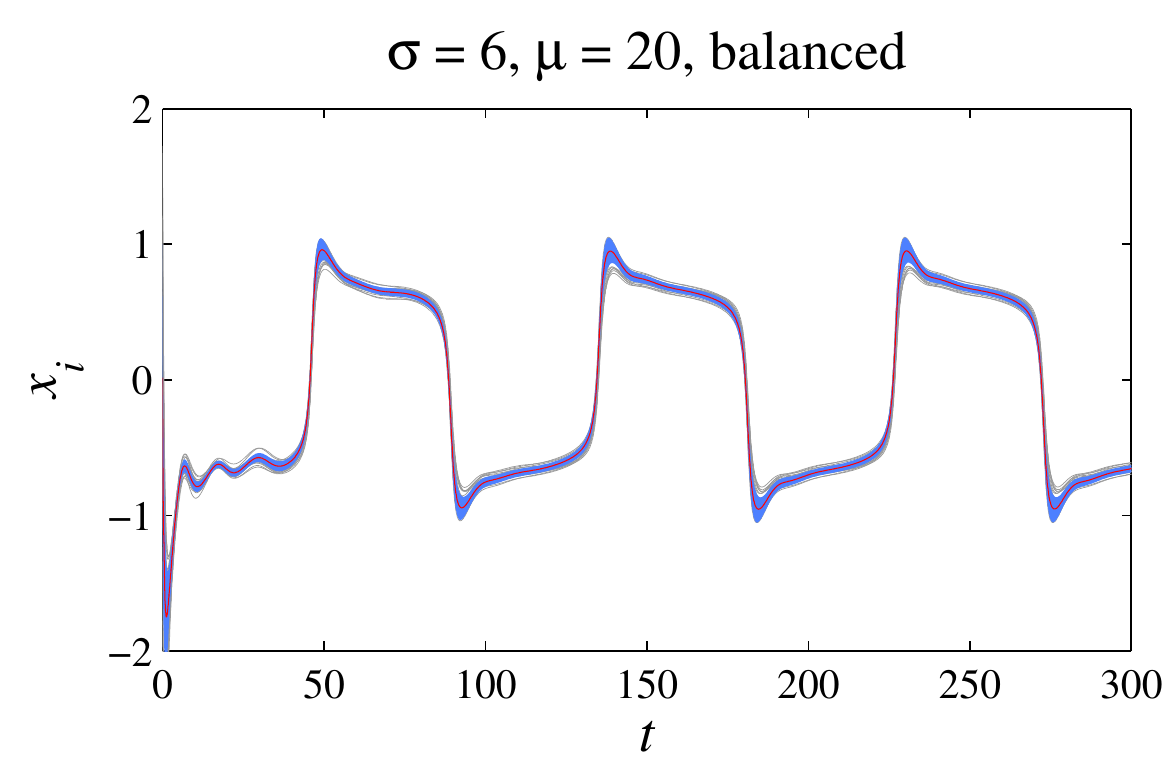}\\
 \hspace{-0.5cm}
 \includegraphics[width=0.33\textwidth]{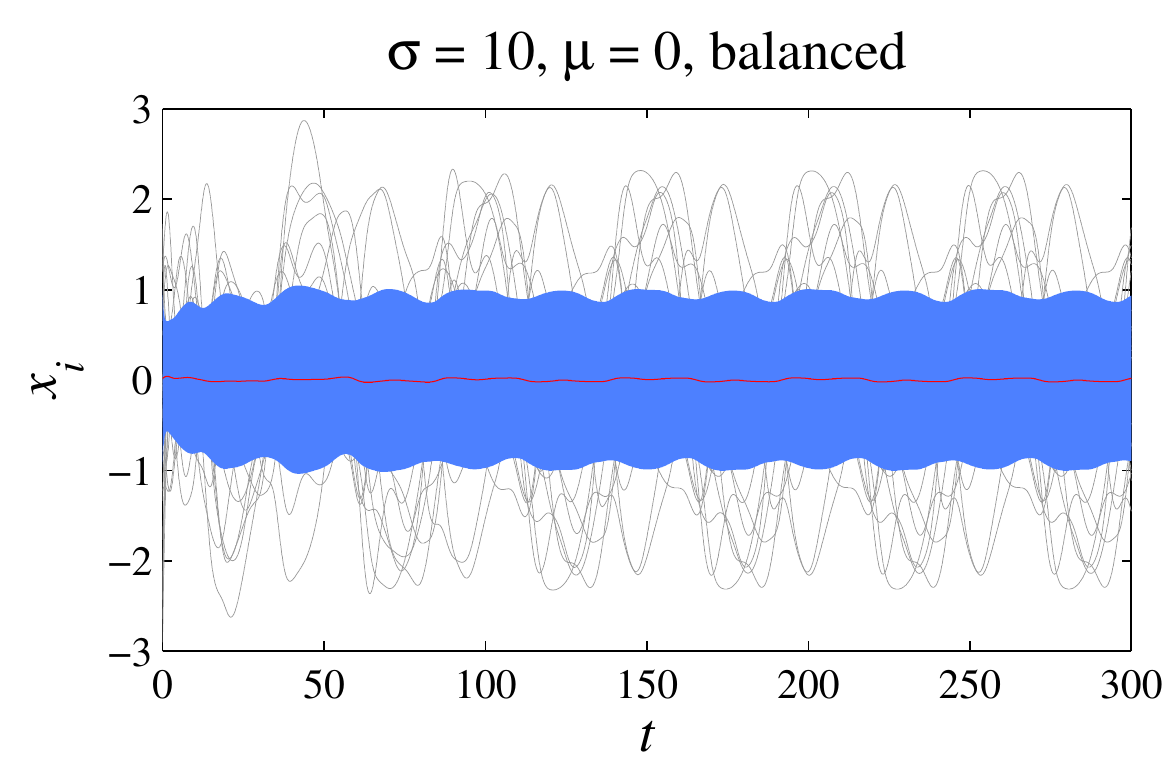}
 \includegraphics[width=0.33\textwidth]{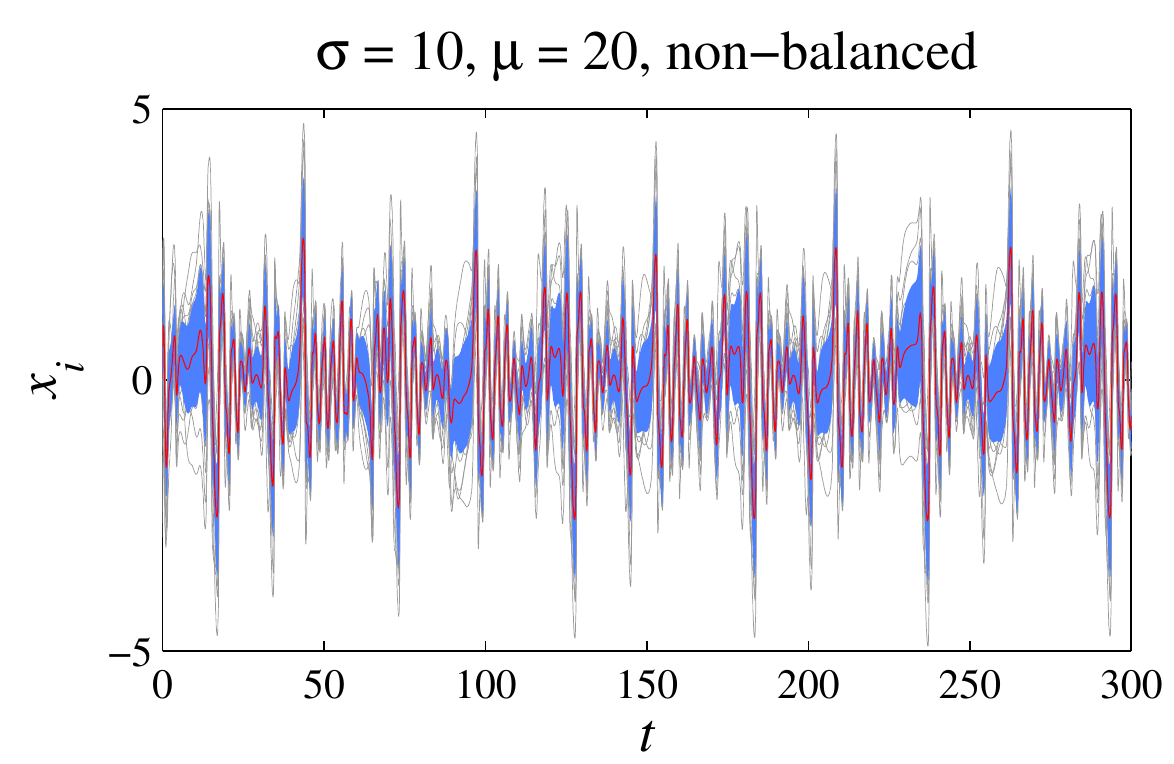}
 \includegraphics[width=0.33\textwidth]{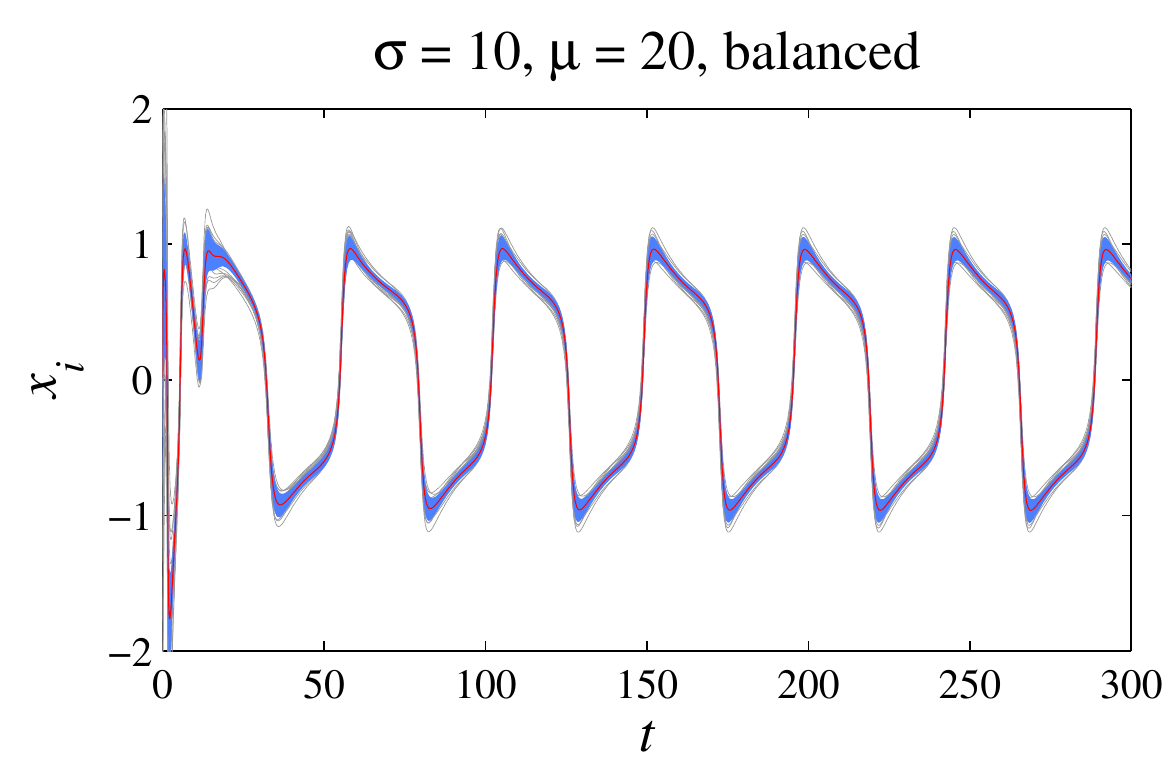}
 \caption{(Color online) Solution of the network equations with $S=\tanh$ and $N=1000$, for different distributions of $\xi$ and $\mu$, and $\chi_j$ uniformly distributed in $[0,1]$. Red curves (front): ensemble average activity, blue shaded areas: average plus minus one standard deviation, gray curves (background): $20$ individual trajectories.  \textbf{Top:} $\xi_{ij}$ are centered independent Gaussian variables (fully connected case) with $m_{i}=\pm 1/\sqrt{n}$. \textbf{Middle:} sparse $\xi$ ($p=0.5$) and \textbf{Bottom:} $\xi$ has a small world structure. In the two last rows, the non-zero coefficients of $\xi$ are uniformly distributed in $[-1,1]$ and $\mathbf{m}$ is a normalized Gaussian vector with zero sum.}
\label{fig:trajectories}
\end{figure*}

\section{NUMERICAL RESULTS}

Let us start by numerically investigating the solutions to~\eqref{eq:GeneralNetwork} for different values of $\mu$, synaptic disorder levels $\sigma$, and also various network topologies (see Fig.~\ref{fig:trajectories}). In the case $\mu=0$ (Fig~\ref{fig:trajectories}, col. 1), the dynamics analyzed in~\cite{sompolinsky-crisanti-etal:88} persists when considering balanced synaptic weights, different $\chi_j$, and sparse or small-world topologies. When taking $\mu>0$ in non-balanced networks, the activity keeps displaying chaotic-like trajectories with irregular dynamics for the mean activity and periods of high synchronization (Fig.~\ref{fig:trajectories}, col. 2). In contrast, this collective behavior becomes highly regular as soon as coefficients are balanced (Fig.~\ref{fig:trajectories}, col. 3) which corresponds to stationary or synchronized oscillatory dynamics. These regular dynamics progressively loose regularity as $\mu$ is reduced or $\sigma$ 
increased. Surprisingly, two matrices with exactly the same eigenvalues (columns 1 and 3 of Fig.~\ref{fig:trajectories}) yield extremely different phenomenology, and therefore spectral analysis is not sufficient. 

\section{REDUCED MODEL}

In order to comprehend the phenomenon of synchronization in balanced networks, we now describe the macroscopic activity through the empirical mean  $z=\frac 1 n \sum_j x_j$ and the individual deviations from the mean $y_i=x_i-z$ (the vector $\y$ is given by $(y_1,\cdots,y_n)$). Expanding $S(\cdot)$ we obtain:
\begin{equation}\label{eq:Mean}
	\dot z = -z + S'(z)\mu \m^t \cdot \y + \varphi(z,\y)+ O(\frac{1}{\sqrt n}).
\end{equation}
The equation on $\y$ simply reads:
\begin{equation}\label{eq:Fluct}
	\dot \y = - \y + \sigma S'(z) \; \xi\cdot \y + \xi\cdot \psi(z,\y)+O(\frac{1}{\sqrt n})
\end{equation}
where $\xi$ is the matrix with elements $\xi_{ij}$, and $\phi$ and $\psi$ correspond to higher order terms in $\y$. In these equations, we observe that the averaged activity is driven by a scalar quantity $\mu\m^t\cdot \y$ which is $\mu$ times the projection of the fluctuations onto the vector $\m$. This quantity is directly related to the standard deviation of $\x$ (which is precisely $\vert \y \vert$) through the formula:
\[\mu\m^t\cdot \y=\mu \vert \y\vert \cos(\theta)\]
where $\theta$ is the angle formed by the vectors $\m$ and $\y$. In order to understand the collective dynamics of the network, we therefore need to express the angle $\theta$ and the dynamics of the standard deviation $\vert \y\vert$. To this end, we need to take a further look to the matrix $\xi$, and in particular its spectrum. Denoting $\lambda_i$ the eigenvalues of $\xi$ corresponding to the normalized eigenvectors $\mathbf{e}_i$, and $c_i=\y^t\cdot \mathbf{e}_i$ the coefficient of $\y$ along the direction of $\mathbf{e}_i$, one can see from equation~\eqref{eq:Fluct} that the coefficients $c_i$ satisfy the equation:
\[\dot{c_i}=(-1+\lambda_i \sigma S'(z))\,c_i + (\xi\cdot \psi(z,\y))^t\cdot \mathbf{e}_i+ O(\frac 1 {\sqrt{n}}).\]
It is therefore clear that the fluctuations are dominated by the modes corresponding to the eigenvalues with largest real part of the matrix $\xi$, which we call \emph{stability exponents}. In order to get a grasp on the dynamics of these processes, let us for a moment neglect the nonlinear terms $\varphi$ and $\psi$. For finite $n$, the system can therefore be in one of two situations: (i) either the stability exponent $\lambda_1$ is real, or (ii) there exist two complex stability exponents $\lambda_1$ and $\lambda_2=\lambda_1^*$.

\begin{figure}[t]
 \centering
\hspace{-5mm} \includegraphics[width=0.5\textwidth]{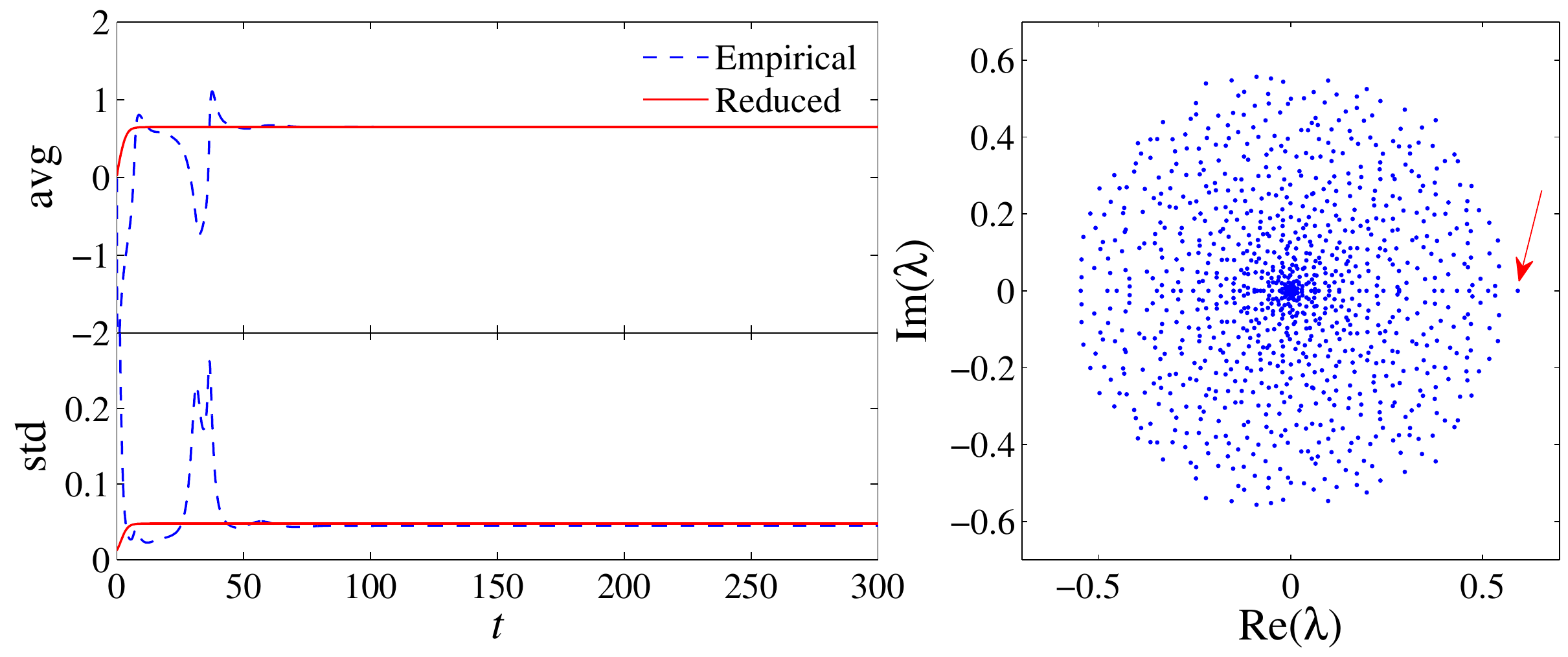}\\
\hspace{-5mm} \includegraphics[width=0.5\textwidth]{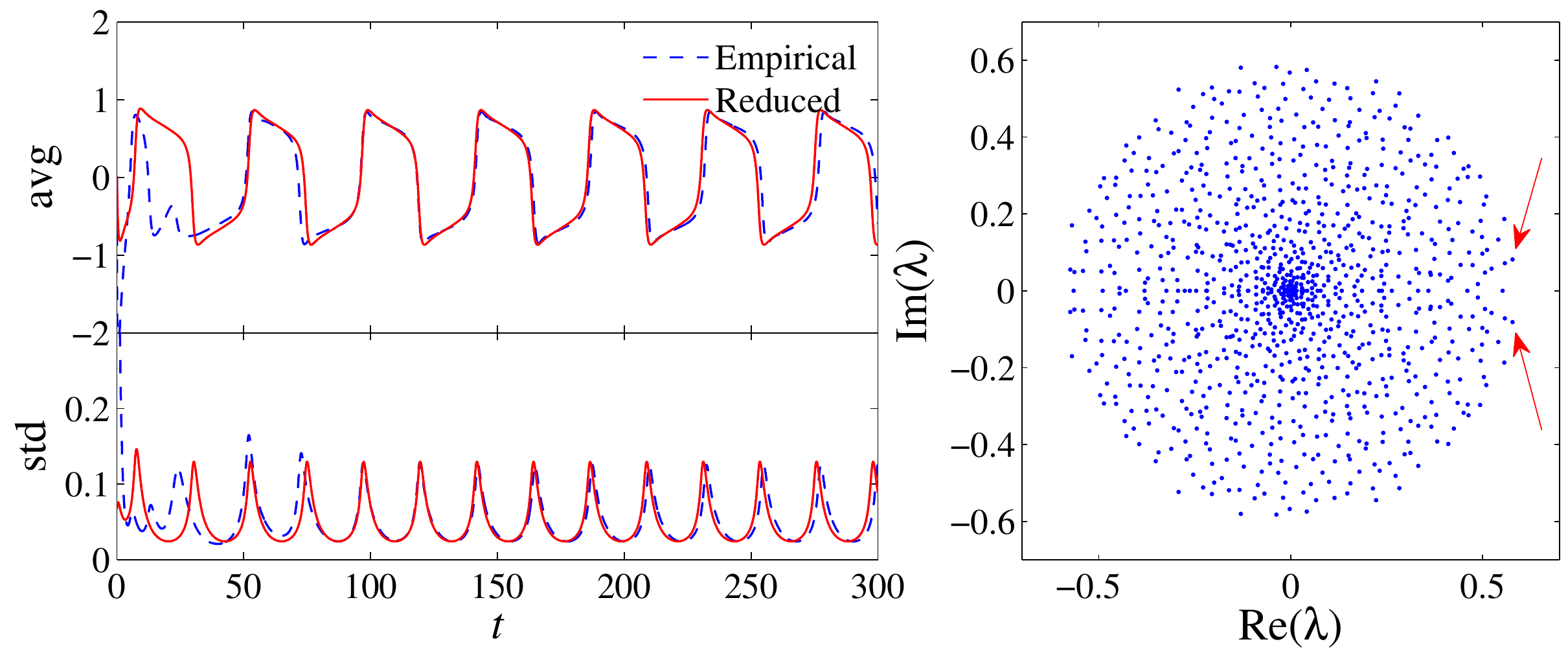}
 \caption{(Color online) Comparison of the empirical average and standard deviation of the original (dotted line) and reduced (solid line) models. \textbf{Top:} Real stability exponent. \textbf{Bottom:}  Complex stability exponents. The right panels show the spectra of $\xi$, red arrows indicate the stability exponents. $\xi_{ij}$ are centered independent Gaussian variables (fully connected case) with $m_{i}=\pm 1/\sqrt{n}$.}\label{fig:reducedvsempirical}
\end{figure}

In situation (i), the fluctuations vector $\y$ will concentrate along the eigenvector $\mathbf{e}_1$ and therefore $\theta$ is the angle formed by $\m$ and $\mathbf{e}_1$ and does not vary in time. In that case, the system reduces, for large $n$, to the equations:
\begin{equation}\label{eq:reducedPitch}
	\begin{cases}
		\dot z &= -z + \mu \cos(\theta) S'(z) \;c_1\\
		\dot c_1 &= (-1+\lambda_1 \sigma S'(z) ) \;c_1.
	\end{cases}
\end{equation}
The origin is always a fixed point of this system, it is the unique solution and it is moreover stable if and only if $\lambda_1\sigma<1$. It looses stability through a pitchfork bifurcation when $\sigma$ exceeds $1/\lambda_1$, and two new fixed points appear. Since $S$ is assumed to be  a sigmoid function, its differential is bell-shaped and therefore is invertible on $[0,1]\mapsto \R^+$, and we denote by $\phi$ this inverse. The fixed points correspond to $z^{\pm}=\pm\phi(\frac{1}{\lambda_1\sigma})$ and $c_1^{\pm}=\frac{\lambda_1\sigma}{\mu\cos(\theta)} z^{\pm}$  and are both stable. In that case, the system will therefore display a transition to a non-trivial equilibrium state when disorder is increased. Figure~\ref{fig:reducedvsempirical} (top panel) displays the solution of both the original and associated reduced systems~\eqref{eq:reducedPitch}. After a transient phase, we see a clear convergence of the empirical mean to $z^\pm$ and of the standard deviation to $c_1^{\pm}/\sqrt{n}$ in agreement with 
the theoretical analysis. 

Situation (ii) is slightly more involved. In that case, the eigenvalue $\lambda_1$ (respectively $\mathbf{e}_1$, $c_1$) is complex and we denote $(\lambda_1^R,\lambda_1^I)$  (resp. $(\mathbf{e}_1^R,\mathbf{e}_1^I)$, $(c_1^R,c_1^I)$) its real and imaginary part. The angle $\theta$ is therefore the angle formed by $\m$ and the direction $\Re(c_1\mathbf{e}_1) = c_1^R \mathbf{e}^R_1-c_1^I \mathbf{e}^I_1$, and may therefore now depend upon time as $c_1$ varies. In these coordinates, the system reduces to the three dimensional ODE:
\begin{equation}\label{eq:ReducedHopf}
	\begin{cases}
		\dot z &= -z + \mu \cos(\theta) S'(z) \vert c_1 \vert\\
		\dot c_1^R &= -c_1^R + \sigma S'(z) (\lambda_1^R c_1^R - \lambda_1^I c_1^I)\\
		\dot c_1^I &= -c_1^I + \sigma S'(z) (\lambda_1^I c_1^R + \lambda_1^R c_1^I)
	\end{cases}
\end{equation}
Here again, the origin is an obvious solution, and the eigenvalues of the Jacobian matrix at this point are $(-1, -1+\sigma\lambda_1^R \pm \mathbf{i} \sigma\lambda_1^I)$. The system undergoes a Hopf bifurcation at $\sigma=1/\lambda_1^R$ with emergence of periodic orbits with frequency $\lambda_1^I/\lambda_1^R$ close to the bifurcation. As disorder is increased, the system will present a transition to periodic dynamics, highly synchronized close to the bifurcation (since the variance of the trajectories, $\vert \y\vert$, is $\vert c_1 \vert/\sqrt{n}$). The system will therefore display a transition to synchronized oscillations as disorder is increased. A typical trajectories are displayed in Fig.~\ref{fig:reducedvsempirical} (bottom panel) and again, after a short transient phase, the original network shows a very good agreement with the periodic orbit of the reduced system~\eqref{eq:ReducedHopf}.

The analysis of this semi-linear approximation also provides a qualitative understanding of the behavior of the original system, at least in a neighborhood of the transition. When $z$ is large enough, $S'(z)$ and $\psi(z,\y)$ are close to $0$ and therefore all components of $\y$ relax exponentially towards ${0}$. This phase of the dynamics contributes to the overall synchronization. As $\y$ decays, the ensemble average $z$ decays as well and $S'(z)$ increases. In the case of real stability exponent, this process leads to a stable stationary state described above. In the case of complex stability exponents, as $z$ approaches, say,  $z^{+}=\phi(1/(\lambda_1^R \sigma))$, the synchronized state $\y=0$ becomes unstable, through a dynamic transition to chaos similar to~\cite{sompolinsky-crisanti-etal:88}. In this phase, due to the complex pair of eigenvalues, the fluctuations $\y$ start to expand following the leading oscillatory modes. The vector $\y$ will eventually cross the plane perpendicular to $\
mmu$, provoking a change of sign of $\cos(\theta)$. The mean activity is now attracted towards negative values until reaching $z^-$. At this point, $\y$ starts decaying again, producing an overshoot of $z$ followed by an attraction towards $z^-$. A symmetrical process then takes place, leading to the emergence of relaxation oscillations. 

\begin{figure}[t]
 \centering
 \includegraphics[width=0.45\textwidth]{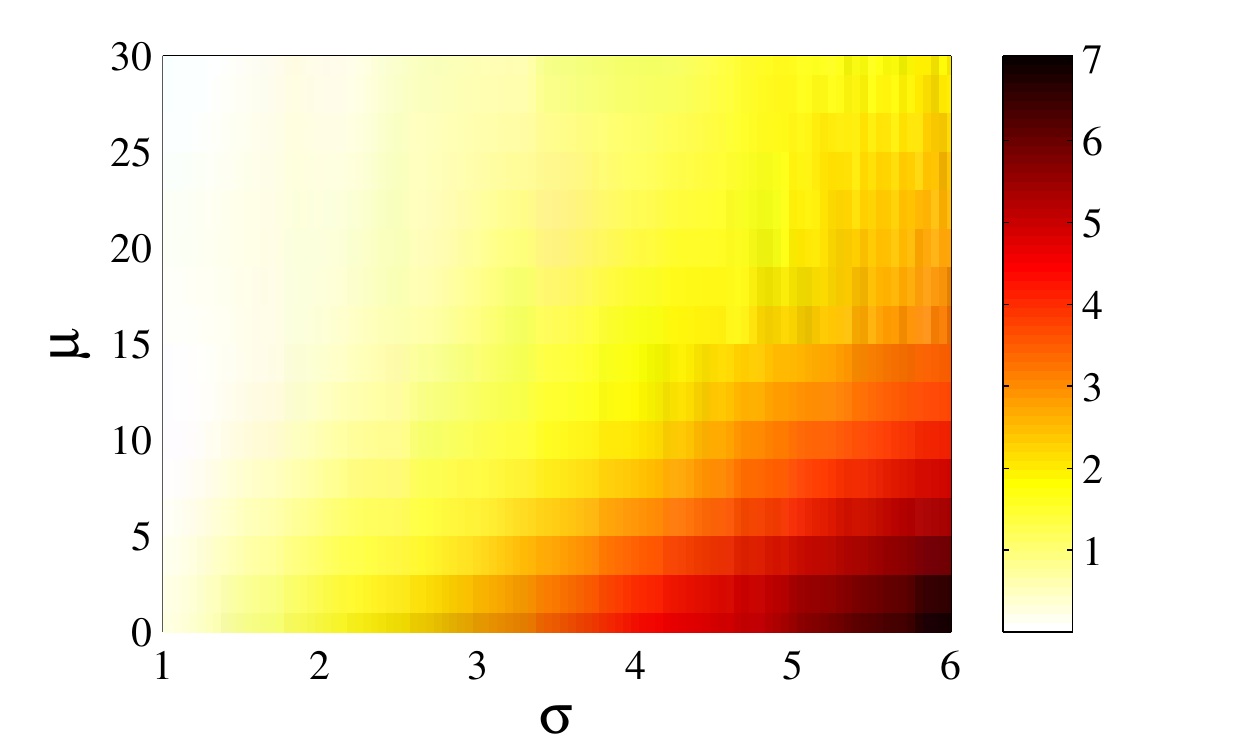} 
 \caption{(Color online) Averaged standard deviation of trajectories of the balanced network after a transient regime, with $N=1000$. $\xi_{ij}$ are centered independent Gaussian variables (fully connected case) with $m_{i}=\pm 1/\sqrt{n}$.}\label{fig:phasediagram}
\end{figure}

We conclude that close to the transition, the system will either show stationary solutions or relaxation oscillations depending only on the nature of the eigenvalue with largest real part of the connectivity matrix. When $\sigma$ is further increased, more complex dynamics appear due to the presence of multiple modes driving $\y$, essentially during the destabilization phase. Although the regularity of the relaxation oscillations is destroyed, the system still display periods of synchronized activity with $z\sim z^{\pm}$, interrupted by short periods of desynchronization. Eventually, when $\sigma$ is much larger than $\mu$, the system is close to a fully disorder network and displays chaotic dynamics as in \cite{sompolinsky-crisanti-etal:88}. The effect of disorder and structure parameters $\sigma$ and $\mu$ is depicted in Fig.~\ref{fig:phasediagram} where we display the averaged standard deviation $\frac 1 T\int_0^T \vert \y(t) \vert dt$ of the trajectories, quantifying the synchronization level of 
the network: increasing $\mu$ reduces the standard deviation of the trajectories, which corresponds to stronger synchronization, whereas increasing $\sigma$ has the opposite effect.



The semi-linear approximation is accurate for the balanced network because once $|\y|$ becomes small it is controlled by $\mu$ and stays small. One can see that this is not the case in the absence of balance in $\xi$. Indeed, assuming that $|\y|$ is small, equation \eqref{eq:Fluct} becomes
\begin{equation*}
 	\dot \y = - \y + S(z)\sum_{j=1}^nJ_{ij}+ \sigma S'(z) \; \xi\cdot \y + \xi\cdot \psi(z,\y)+O(\frac{1}{\sqrt n})\ .
\end{equation*}
The second term in the r.h.s., which vanishes in the balanced scenario, is now of $O(1)$ forcing $|\y|$ to grow. This accounts for the key difference between the balanced and non-balanced networks, namely that in the latter synchronized regimes cannot persist in time.

\section{DISCUSSION}

In contrast to usual non-balanced mean-field theory, the activity of balanced networks dramatically depends on the properties of extremal values of the connectivity matrices rather than on macroscopic estimates, and therefore remains random in the large $n$ regime. This motivates the study of the nature of the stability exponents \footnote{In~\cite{rider-sinclair:12} the authors recently considered this question in the case of the real Ginibre ensemble but universal laws are still lacking.}. Numerical investigations tend to show that, as $n$ increases, the probability of the stability exponent being complex increase and its imaginary part decreases. In other words, for larger systems, the probability of having slow periodic relaxation oscillations becomes larger, and the period of these oscillations shorter.  

This study can provide some insight on the biological functionality of excitatory-inhibitory balanced networks. 
Balanced connectivity accounts for a number of fundamental biological phenomena such as maintaining a dynamic range in the face of massive synaptic input~\cite{shadlen-newsome:94}, presenting a rich repertoire of behaviors~\cite{brunel:00, vanvreeswijk-sompolinsky:96}, and has been proposed as an explanation for the selectivity to orientations in non-structured cortical areas~\cite{hansel-vanvreeswijk:12}. Here we show that it can also play a fundamental role in the emergence of synchrony and regular oscillatory dynamics. 
We also show that disorder plays a crucial role on the dynamics of the network. Experimental results have shown that it significantly impacts the input-output function, rhythmicity and synchrony of neuronal networks~\cite{aradi-soltesz:02,santhakumar2004plasticity,soltesz2005diversity}. These studies relate disorder to transitions between physiological and pathological behaviors, suggesting that specific levels of disorder favor synchronization of neuronal networks. Our work provides a theoretical approach to this question.
In conclusion, we have identified two key connectivity parameters $\mu$ and $\sigma$, which may be interesting to measure experimentally and may be involved in the regulatory processes controlling neuronal activity.

A new form of periodic synchronization therefore arises in balanced networks. This surprising behavior is due to the interaction between the structure and the disorder present in the connectivity. It is remarkable that the extremely regular and yet non-trivial macroscopic dynamics are driven by the chaotic fluctuations of $\y$ rather than being driven by the ensemble average itself. Moreover, the transition we exhibit enjoys relatively broad universality. Indeed, our developments did not rely on a specific structure of $\xi$ or $\m$, beyond balance condition and appropriate scaling. This means that, as shown in figure~\ref{fig:trajectories}, the results hold for general distributions of $\xi$ and $\m$, such as sparse or small world-type connectivity matrices. This phenomenon is also a novelty in the sense that many systems that exhibit synchronized oscillations are ensembles of coupled oscillators or excitable systems~\cite{complex-networks} whereas here individual elements are not natural oscillator:
  both the synchronization and periodicity are emerging properties of the coupling. This work therefore opens the way to a more detailed understanding of the dynamics of random balanced networks, and shows that this class of model displays novel properties, which can be explained through the analysis of reduced low-dimensional models.

\bibliographystyle{apsrev4-1}    
\bibliography{PhysRev}

\end{document}